# A review of modeling applications using ROMS model and COAWST system in the Adriatic sea region

Sandro Carniel,
*Senior Research Scientist, ISMAR-CNR, Castello 2737-F, Venezia 30122, Italy. E-mail: sandro.carniel@ismar.cnr.it*

Aniello Russo,
*Assistant Professor, Dept. of Life and Environmental Sciences, Università Politecnica delle Marche, Ancona 60131, Italy. Email: a.russo@univpm.it*

Alvise Benetazzo
*Research Scientist, ISMAR-CNR, Castello 2737-F, Venezia 30122, Italy. E-mail: alvise.benetazzo@ve.ismar.cnr.it*

**ABSTRACT**
From the first implementation in its purely hydrodynamic configuration, to the last configuration under the Coupled Ocean-Atmosphere-Wave-Sediment Transport (COAWST) system, several specific modelling applications of the Regional Ocean Modelling Systems (ROMS, www.myroms.org) have been put forward within the Adriatic Sea (Italy) region. Covering now a wide range of spatial and temporal scales, they developed in a growing number of fields supporting Integrated Coastal Zone Management (ICZM) and Marine Spatial Planning (MSP) activities in this semi-enclosed sea of paramount importance including the Gulf of Venice.
Presently, a ROMS operational implementation provides every day hydrodynamic and sea level 3-days forecasts, while a second one models the most relevant biogeochemical properties, and a third one (two-way coupled with the Simulating Waves Nearshore (SWAN) model) deals with extreme waves forecast.
Such operational models provide support to civil and environmental protection activities (e.g., driving sub-models for oil-spill dispersion, storm surge, coastal morphodynamic changes during storms, saline wedge intrusion along Po River), in a growing context of stake-holders at regional, national and international level.
Besides, ROMS and COAWST research based activities are also carried out, mostly aiming at investigating sediment transport, eggs and larvae dispersion, hypoxic events in the basin; through successive nesting very high resolutions nearshore the Italian coast can be reached, allowing to simulate river mouth environments and artificial reefs.
Resulting outputs, written in NetCDF CF compliant format, are delivered via THREDDS Data Server to a growing number of users around the world.

**Introduction**
Delimited on three of its sides by the Italian coasts and the Balkans, the Adriatic Sea is a NW-SE elongated semi-enclosed basin, about 700 km long and 200 km wide, opening to the Mediterranean Sea through the Otranto Strait. It is characterized by so-called Bora (from NE) and Sirocco (from SE) winds, which can cause high waves and littoral erosion mostly in the NW shallow coasts (Signell et al., 2005; Martucci et al., 2010; Bignami et al., 2007). The Adriatic Sea is an area of great environmental and socio-economic value (e.g. the gulf of Venice); thanks to the easiness of access and to the number of scientific Institutions based along its coast, it has very often been selected as test case study for international, EU and national funded projects. Its northern part, very shallow and collecting a large river runoff (with drainage basin characterized by intensive farming and hosting several industries), hosts relevant touristic, fishery, extracting and maritime transport activities. This area is presenting also several environmental concerns, like eutrophication, anoxic episodes, massive mucilaginous appearance, harmful algal blooms, small pelagic fishes stock depletion, pollution.
We focus here on some modeling efforts carried out using the Regional Ocean Modeling System (ROMS, www.myroms.org) family, a community, three-dimensional, hydrostatic, finite difference model resolving the Reynolds Averaged Navier Stokes equations. From the first Adriatic Sea implementation in its purely hydrodynamic configuration that traces back to more than ten years ago (Carniel et al., 2009), to the currently available Coupled Ocean-Atmosphere-Wave-Sediment Transport (COAWST) system, including wave-current-sediment interactions, several specific applications covering a wide range of spatial and temporal scales have been put forward, developed in a growing number of fields. Here we review some of these implementations, based on recent code releases and resolution grids.
Besides operational implementations, ROMS and COAWST systems are employed for cutting-edge research activities, mostly aiming at investigating sediment transport (Sclavo et al., 2013), coastal dynamics and environmental aspects, eggs and larvae dispersion, hypoxic events etc. Moreover, coupled applications among ocean-atmosphere-wave-sediment and the possibility of employing both one-way and two-way successive nesting techniques allowed to reach very high resolutions nearshore the Italian coast, and to simulating also complex regions such as small river mouth environments and artificial reefs. Resulting outputs, written in NetCDF CF compliant format, are delivered via THREDDS Data Server to a growing number of users around the world (Bergamasco et al., 2012).

**1. AdriaROMS 4.0 system, hydrodynamic forecast and dedicated sub-models**

AdriaROMS 4.0 is a system running on a regularly spaced 2 km resolution grid of the whole Adriatic sea, providing numerical output forecasts every day. Air-sea heat, momentum and water fluxes are interactively computed from the COSMO-I7 hourly outputs. Main tidal components are imposed at the open boundary, Po river discharge is introduced using real-time data, whereas runoff of other 48 rivers and karstic springs are introduced by using monthly climatologies derived from the literature. Temperature, salinity, sea level and velocity conditions at the open-boundary (Otranto Strait) are provided by the Italian Operational Oceanography Group's (GNOO) Mediterranean Forecasting System.

AdriaROMS 4.0 outputs are used at ARPA-SIMC to feed several operational applications for purposes of civil and environmental protection. A morphodymanic 1D model is applied to coastal profiles in some exposed areas of the Emilia Romagna, south of the Po River delta, for early warning in case of sea inundation; it is forced at the boundary of the profiles by the total water level derived by AdriaROMS 4.0 outputs and by the significant wave height obtained by SWAN-EMR (Valentini et al., 2007).

The total water level from AdriaROMS 4.0, together with 10 m wind computed by COSMO-I7 and the significant wave height by SWAN-EMR, are also used as boundary conditions for the assessment and the prediction of the bathing water quality in front of Rimini, a famous beach resort of the NW Adriatic coast. For this application, the specific model is based on Delf3D (Deltares, 2010) and it has been implemented within the regional project of Emilia-Romagna Region "Previbalneazione". The model has been implemented and used for the definition of the bathing water profiles and management, according to the European Union Directive 2006/7/CE.

Other applications adopt AdriaROMS 4.0 outputs, albeit not in a strictly operational configuration, deal with Rapid Response forecasts of oil spill dispersion in case of release at sea or from the coast (in conjunction with regional and Coast Guard authorities) and saline wedge intrusion predictions within the Po river.

**2. EMMA system, hypoxia events forecast**

Due to its peculiar morphologic, oceanographic and climatic conditions, together with anthropic pressure factors, the Adriatic sea is extremely propitious to the growth of hypo-anoxic (low or no oxygen) events in waters. Anoxia determinates consistent damages in terms of environmental quality of the coast, quality and quantity of ichthyic and mariculture products. Consequently, the innovative monitoring project EMMA (Environmental Management through monitoring and Modelling of Anoxia, http://emma.bo.ismar.cnr.it/) was designed, under LIFE-Enviroment framework, including Italian and Slovenian partners. The project key aspect is the integration of different actors from academy, public authority and fishing association, encouraging a synergy between research and business at national and European level, in the direction of a MSP approach.

EMMA is an operational implementation having as main purpose the short term forecasting of hypoxic events in the northern Adriatic Sea, with particular attention to the Rimini area, as described in Russo et al. (2009). The modeling system is ROMS, adopting the Fennel biogeochemical module available within ROMS, and reproducing pelagic nitrogen cycle processes in the water column and remineralisation processes at the water-sediment interface. The Fennel module includes inorganic nitrogen (nitrate and ammonium), phytoplankton and zooplankton biomass, small (<10 mm) and large detritus, all expressed in terms of nitrogen and carbon concentrations, inorganic carbon and dissolved oxygen dynamics. Po river discharge and other rivers have been included, adding also the climatologic inorganic nitrogen load. Air-sea fluxes are computed using outputs produced by a non-hydrostatic limited-area atmospheric model, COSMO-I7, Italian operational implementations of the Consortium for Small-scale Modelling (COSMO, www.cosmo-model.org) model, with horizontal resolution of 7 km, boundary conditions provided by the ECMWF and assimilating available observations.

The integration time covers the period since late spring 2007; horizontal resolution is ~2 km and outputs are provided every 3 hours. Details of the implementation can be found in Russo et al. (2009) and Russo et al. (2013).

During 2012 EMMA system results (namely, velocity and density fields) were provided in RT (Real Time) to the CNR-ISMAR R/V *Urania* during the sea-truth campaign "Operation Dense Water" (ODW). The fields received on board allowed to provide a striking example of "adaptive sampling strategy" that maximized the probability to detect, using CTDs, L-ADCP and XBTs, dense water masses flowing to the southern Adriatic sea after having been formed in the northern basin during one of the most severe winter of the last century.

### 3. NA-COAWST (coupled wave-current system)

Resourcing to several test cases described in Benetazzo et al. (2013), an operational version of COAWST (hereafter NA-COAWST) in the northern Adriatic basin has been setup, including two-way wave-current interactions; the coupling between ROMS and SWAN models is performed though the Modeling Coupling Toolkit. NA-COAWST domain covers the northern Adriatic sub-basin at 0.5 km horizontal resolution, in a shallow (max 100 m) domain, the grid having been designed to fit exactly into the AdriaROMS 4.0 one. AdriaROMS 4.0 and SWAN ITALIA (Valentini et al., 2007) operational models provide conditions for currents, level, temperature, salinity and wave characteristics at the southeastern open boundary, where tidal components are also imposed. NA-COASWT has been running in operational mode since November 25, 2011.

In the configuration adopted, the ocean model provides the wave model with currents and free surface elevation using the formulations thoroughly discussed in Benetazzo et al. (2013).

The availability of NA-COAWST outputs is relevant to the coastal stakeholders of the region in the direction of ICZM and of a careful planning of this marine area. Indeed, while for most of the sea conditions the forecast of the pure currents is usually sufficient, during severe sea storms the coupled wave-current model allows a more careful determination of small-scale features, including regions larger waves and sediment blobs detection. This is of particular interest also to ship-routing, fishermen and industrial activity in the Adriatic sea. First results from the sensors of the CNR-ISMAR tower "*Acqua Alta*" (see www.ismar.cnr.it) and from the acoustic Doppler current profiler (AWAC) deployed in winter 2013 in the area of Jesolo in the northern Adriatic Sea, seem indeed to support the fact that NA-COAWST system is providing better forecasts, that have been promptly adopted also in the context of locally existing MSP initiative providing hazards such as the significant wave height to Decision Support Systems.

### 4. Transport and dispersion of small pelagic fish eggs and larvae

Increasing attention has been recently given to studies concerning transport and diffusion of small pelagic fishes (anchovies and sardines) eggs and larvae using an Individual Based Model (IBM) driven by ROMS outputs (Russo et al., 2013). A series of IBM simulations were performed, covering two years with different atmospheric and marine characteristics (in terms of current fields and river runoff).

### 5. High-resolution nearshore sediment dynamics (Bevano river)

In order to show how integrated numerical tools are becoming suitable both for preliminary investigations and for planning effective littoral management as well, Carniel et al. (2012) and Ciavola et al. (2012) described the implementation of the integrated wave-currents-sediment numerical model consisting of ROMS, in its fully 3-D, two-way coupled version with the wave model SWAN and again the dedicated sediment transport module (Carniel et al., 2011).

The system was setup in the Bevano river region, an area of the northern Adriatic Sea characterized by a microtidal and low energy wave environment. In 2006 the river mouth was artificially modified to prevent dune erosion and decrease river flooding, by means of timber engineering works. Local authorities needed to understand and model the morphological changes induced by likely extreme hydro-meteorological events. A series of hydrodynamic and morphological scenarios by means of a coupled numerical model were then tested, using data from field surveys.

The probability of a second inlet opening, either by river breaching or overwashing, was another topic regarded as important by the stakeholders of the region in the direction of ICZM and of a careful planning of this marine area.

While for most of the sea conditions the current inlet has been shown to be the main source of water escape to the sea, results indicated that a secondary inlet can actually be opened in case of a 30 year flood occurrence; moreover, the modeling results confirmed the dominance of the ebb tide in creating a small ebb delta/swash bar. These indications were then adopted in a context of local MSP and management, leading to a "minimum maintenance option strategy", supported by regular repairs of the timber structures.

### Conclusions

The paper has briefly recalled and described some applications of ROMS-COAWST modeling systems that have been turning out to be supportive to ICZM and MSP activities in the Adriatic sea. The discussed studies briefly described operational applications, including also hindcast studies oriented at the improvement of the understanding of the basic processes of this semi-enclosed sea of paramount importance.

The numerical tools outlined are now providing support to scientific, civil and environmental protection applications (e.g., driving sub models for prediction of oil-spill dispersion, storm surge, coastal morphodynamic changes during storms, saline wedge intrusion along Po river), in a growing context of stake-holders at regional, national and international level. The operational implementations providing assistance and guidance in the field of ICZM are benefiting from findings developed within research based activities, focusing on the investigation of coastal dynamics, sediment transport, general circulation, hypoxic events, eggs and larvae dispersion. All together, these case studies could demonstrate the suitability of employing open source, community numerical models for complex applications, embedding several state-of-the-art capabilities (among which we recall an up-to-date bottom boundary layer description, wetting and drying capabilities, advanced vertical mixing and wave-current interactions schemes (Kantha and Carniel, 2003; Carniel et al., 2009), a biogeochemical module). Some recent features, such as the possibility of using both one-way and two-way successive nesting and full two-way coupling, have increased the versatility of COAWST tool, allowing to reach very high resolutions nearshore the Italian coast and to simulate with extreme detail river mouth environments.

When numerical models are brought in direct contact with ICZM or MSP activities, one of the most striking problems that exists is related to the so-called "Users and Data Producers entry barriers". Great strides have been made by the Adriatic ROMS-COAWST community to overcome the usually existing bottleneck in the distribution and use of the modelled ocean data obtained by the operational efforts described (Signell et al., 2008; Bergamasco et al., 2012). More specifically, to allow *data providers* (e.g. numerical modelers producing *big data*) to serve their data with no need of modification via standard web services, some of the results of the above described ROMS and COAWST systems are written in NetCDF format (compliant to the Climate & Forecast convention) and distributed via a THREDDS Data Server (TDS). Since the TDS metadata service can be harvested by catalog brokering systems, data users can easily access the standardized data by using standardized queries for data discovery (e.g. OpenSearch, OGC Catalog Services, etc.). After deciding which time/space portion has to be transferred, they can select among a variety of tools, including 3D open-source viewers, to promptly visualize or carefully examine them. The brokering approach to harvest metadata from many different services and to read data from many different formats into common data models greatly improves model access and interoperability, unlocking information from other fields (e.g., social and economic studies). These are very desirable properties in the direction of a "INSPIRE compliant web service" (see also http://inspire.jrc.ec.europa.eu). Some examples can be assessed at site http://tds.ve.ismar.cnr.it:8080/thredds/catalog.html.

**ACKNOWLEDGEMENTS**

We gratefully acknowledge the support from the National Flagship Project "RITMARE"(SP3-WP4-A2 and SP3-WP4-A3) funded by MIUR. The operational implementation of AdriaROMS 4.0, NA-COAWST and XBeach at ARPA-SIMC has been partially funded by the EU VII Framework Program through the MICORE project (Grant Agreement No. 202798). The operational implementation of EMMA forecasting system at DISVA has been partially funded by the EU LIFE program (contract LIFE04 ENV/IT/000479). The development of COAWST system in the Adriatic Sea has been partially supported by funding from the EC FP7/2007-2013 under grant agreement n° 242284 (Project "FIELD_AC") and by the FIRB RBFR08D825 grant (Project "DECALOGO").